\documentclass[10pt, conference, compsocconf]{IEEEtran}
\usepackage[utf8]{inputenc}
\usepackage[T1]{fontenc}
\usepackage[table]{xcolor}
\usepackage{graphicx,tabularx,url,amsmath,amssymb,float,tikz,siunitx,soul,verbatim,dsfont,subfigure}
\DeclareMathAlphabet{\mathpzc}{OT1}{pzc}{m}{it}
\usepackage[lined,ruled,linesnumbered]{algorithm2e}
\usepackage[pdftex, %
 			pdftitle={title}, %
 			pdfcenterwindow=true, %
 			pdfstartview=Fit, %
 			pdfdisplaydoctitle=true, %
 			pdfauthor={authors}, %
 			pdfpagemode=UseNone,
 			bookmarksopen=true,
 			breaklinks=false,
 			bookmarksnumbered=true,
 			colorlinks=true,
 			urlcolor=blue,
 			anchorcolor=black,
 			citecolor=black,
 			linkcolor=black,
 			pdfborder={0 0 0}]{hyperref} 
\newtheorem{definition}{Definition}

\newcommand{\tcompany}{\mbox{the Company}}

\newcommand{\company}{\mbox{Company}}
\newcommand{\tuser}{\mbox{the User}}

\newcommand{\user}{\mbox{User}}

\newcommand{\li}{\hat{l}_i}

\newcommand{\hlc}[2][yellow]{ {\sethlcolor{#1} \hl{#2}}}
\newcommand{\cmanos}[1]{\hl{[\textbf{Manos:} #1]}}

\newcommand{\caron}[1]{\hlc[green]{[\textbf{Aron:} #1]}}

\newcommand{\offserv}{\ensuremath{S}}

\newcommand{\expserv}{\ensuremath{\sigma(\connectiontime,\offserv)}}

\newcommand{\mc}{\mathcal}
\newcommand{\mf}{\mathbf}
\newcommand{\privacy}{p}

\newcommand{\strategyset}{\mathcal{S}}
\newcommand{\mixedstratcomp}{\mathbf{\Phi}}

\renewcommand{\vec}[1]{\ensuremath{\boldsymbol{#1}}}

\newcommand{\usertypes}{\mathcal{A}}

\newcommand{\connectiontime}{\ensuremath{t}}

\newcommand{\totaltime}{T}

\newcommand{\dfn}{:=}
\newcommand{\loc}{\vec{l}}

\newcommand{\estloc}{\ensuremath{\loc_{est}(\tau)}}

\newcommand{\privacyRandomVariable}{\Pi}

\newcommand{\ci}{\mbox{CI}}
\newcommand{\distance}{\left \Vert \loc(\tau) - \estloc \right \Vert}
\newcommand{\newdistance}{\left \Vert \loc - \li \right \Vert}
\newcommand{\comppayoff}{\Psi \, t_i - \Theta \, \sum\limits_{j\in \mc{S}_C}\phi_j \, S_j}
\DeclareMathOperator*{\E}{\mathbb{E}}

\newcolumntype{C}[1]{>{\centering\arraybackslash}p{#1}}
\newcolumntype{L}[1]{>{\arraybackslash}p{#1}}
\definecolor{Gray}{gray}{0.88}
\newtheorem{theorem}{Theorem}
\newtheorem{lemma}{Lemma}

\ifCLASSINFOpdf
\else
\fi
\newcommand\copyrighttext{%
\footnotesize 
Copyright by IEEE.~Personal use of this material is permitted.~However, permission to reprint/republish this material for advertising or promotional purposes or for creating new collective works for resale or redistribution to servers or lists, or to reuse any copyrighted component of this work in other works must be obtained from the IEEE.\\
Panaousis, E.; Laszka, A.; Pohl, J.; Noack, A.; Alpcan, T., ``Game-Theoretic Model of Incentivizing Privacy-Aware Users to Consent to Location Tracking,'' in Trustcom/BigDataSE/ISPA, 2015 IEEE , vol.1, no., pp.1006-1013, 20-22 Aug. 2015, doi: 10.1109/Trustcom.2015.476
URL: \url{http://ieeexplore.ieee.org/stamp/stamp.jsp?tp=\&arnumber=7345384\&isnumber=7345233}}
  
\newcommand\copyrightnotice{%
\begin{tikzpicture}[remember picture,overlay]
\node[anchor=south,yshift=10pt] at (current page.south) {\fbox{\parbox{\dimexpr\textwidth-\fboxsep-\fboxrule\relax}{\copyrighttext}}};
\end{tikzpicture}%
}

\begin{document}
\title{Game-Theoretic Model of Incentivizing Privacy-Aware Users to Consent \\ to Location Tracking}
\author{
\IEEEauthorblockN{
Emmanouil Panaousis\IEEEauthorrefmark{1},
Aron Laszka\IEEEauthorrefmark{2},
Johannes Pohl\IEEEauthorrefmark{3},\\
Andreas Noack\IEEEauthorrefmark{3}, 
and Tansu Alpcan\IEEEauthorrefmark{4}
}

\IEEEauthorblockA{
\IEEEauthorrefmark{1}
University of Brighton, UK}

\IEEEauthorblockA{
\IEEEauthorrefmark{2}
Institute for Software Integrated Systems, Vanderbilt University, Nashville, USA}

\IEEEauthorblockA{
\IEEEauthorrefmark{3}
University of Applied Sciences Stralsund, Germany}

\IEEEauthorblockA{
\IEEEauthorrefmark{4}
University of Melbourne, Australia}
}

\maketitle
\pagenumbering{gobble}.
\copyrightnotice

\begin{abstract}
Nowadays, mobile users have a vast number of applications and services at their disposal. Each of these might impose some privacy threats on users' ``Personally Identifiable Information'' (PII). Location privacy is a crucial part of PII, and as such, privacy-aware users wish to maximize it. This privacy can be, for instance, threatened by a company, which collects users' traces and shares them with third parties.~To maximize their location privacy, users can decide to get offline so that the company cannot localize their devices. The longer a user stays connected to a network, the more services he might receive, but his location privacy decreases. In this paper, we analyze the trade-off between location privacy, the level of services that a user experiences, and the profit of the company. To this end, we formulate a Stackelberg Bayesian game between the User (follower) and the Company (leader). We present theoretical results characterizing the equilibria of the game. To the best of our knowledge, our work is the first to model the economically rational decision-making of the service provider (i.e., the Company) in conjunction with the rational decision-making of users who wish to protect their location privacy. To evaluate the performance of our approach, we have used real-data from a testbed, and we have also shown that the game-theoretic strategy of the Company outperforms non-strategic methods. Finally, we have considered different User privacy types, and have determined the service level that incentivizes the User to stay connected as long as possible.
\end{abstract}

\thispagestyle{plain}
\pagestyle{plain}
\section{Introduction}
\label{sec_intro}
The prevalence of smartphones brings to end users not only new applications and services but also privacy risks. These risks are due to the possible disclosure of vast amount of private information. In this paper, we investigate how location privacy is affected by the amount of time a User is connected to a wireless local area network (WLAN). We propose a game-theoretic model to capture the interaction between a Company and a User. The former offers some services to the latter, while he is connected to a WLAN that belongs to the Company. We assume that the Company uses a wireless communication technology to localize users in order to increase its profits by launching  targeted advertisements or by selling User location data to third parties. It is worth noting here that our  analysis is not restricted to localization within a WLAN. It can, for instance, be rectified to increase location privacy in scenarios where phones can be tracked without using their GPS or WiFi data, e.g.\,by studying only their power usage over time, as in~\cite{powerspy}.
 
Our work is motivated by the observation that location disclosure entails different privacy risks, and we can realistically say that the location data is valuable to the Company. Suppose, for example, that the Company has established its wireless network within a shopping centre. The location data of the visitors can be utilized for:
\begin{itemize}
 \item \emph{optimization of stores:} $\tcompany$ can optimize the store design based on heat-maps of customer movements; 
 \item \emph{targeted advertisements:} if $\tcompany$ knows the location of customers, it can send product information based on their location, creating \emph{location-based spam}; 
 \item \emph{profiling:} from $\tuser$'s long-term location information, $\tcompany$ can create profiles, and use them for strategic decisions, or even sell this information to third parties.
\end{itemize}

In order to obtain the desired location data, the Company establishes a passive localization system based on signal information (e.g.,\,Received Signal Strength (RSS)) of the users' devices. During connection time, the User can be localized and therefore the more the User stays connected, the more location traces can be collected by the Company. The latter offers services to the User, which can compensate the location privacy loss. These services may include free broadband access, geolocation services, and discounts for certain products or lotteries.

This paper is organized as follows. The system model, including both the Company and the User, is described in Section \ref{sec_model}. In Section \ref{sec_lpg}, we formulate the Location Privacy Game (LPG) by defining the players' strategies, types, and payoffs. Section~\ref{sec_analysis} is dedicated to the theoretical analysis presenting the equilibria conditions of the game, and deriving the User's best response and the Company's optimal strategy in LPG. In Section \ref{sec_case_study}, we present the performance evaluation results, which demonstrate the effectiveness of our game-theoretic approach. The related work is discussed in Section \ref{sec_related_work}, while Section \ref{sec_conclusion} concludes the paper.

\section{System Model}
\label{sec_model}
In our model, we assume a $\company$ which controls the communication infrastructure ($\ci$) (e.g.,\,WiFi network) of a building (e.g.,\,a shopping centre) and offers services to the visitors when they are connected to CI. We consider the User as the entity that can utilize these services, and at the same time, he can be located by $\tcompany$, which leads to suffering some location privacy loss. For a list of symbols used in this paper, see Table \ref{tab_symbols}.

\subsection{Passive Localization System}
We assume that $\tcompany$ maintains a passive indoor localization system to determine the location $\loc(\tau)$ of the User at time $\tau$.
The passive localization system determines a location estimate $\estloc$, which is an approximation of the User's true location at time $\tau$. The precision of this approximation is determined by the number of data packets that the User transmits per second, i.e., the more data the User sends the more precise $\estloc$ becomes; however, modeling this relationship is out of the scope of this paper.  This approach is different from cases where the User actively reports his location in order to use LBSs~\cite{manshaei_game_2013, shokri_hiding_2013}, as we assume that localization occurs without the User's active participation.

Any position estimate $\estloc$ is biased with an error
\begin{equation}
  l_{err}(\tau) \dfn \distance.
\end{equation}
As the User is moving, $l_{err}(\tau)$ can take different values (i.e., $l_{err}(\tau)$ is a random variable). We denote the expected value $\E[l_{err}(\tau)]$ by $\hat{l}$.

\subsection{Location Privacy}
We assume that the User is roaming within the Company's area for time $T \in \mathbb{Z}$, but his device stays connected to the CI of the Company only for time $t \in \mathbb{R}: \delta \leq t \leq T$, where $\delta$ is very small value. We have assumed here that every User needs some minimal amount of connection time $\delta$, for example, in order to become aware of the services that the Company offers. The lower the value of $t$, the lower the location privacy loss of the User, as the User can be located only during $t$, since there are no data packets transmitted when the User is not connected. Then, the Users' location privacy, when connected to CI for time $t$, equals
\begin{equation}
\label{eq:privacy}
 p(t) \dfn \frac{T}{t} \, \hat{l} . 
\end{equation}
\begin{samepage}
 In order to increase his location privacy, the User seeks a minimum $t$ with respect to some minimum required service level, which will define later in this section. This is based on the assumption that the longer the User stays connected, the higher level of service he receives. 

\subsection{User Types}
In this paper, we assume that there are multiple User types. This is motivated by real-world scenarios where a company provides some services and several users (i.e., of different types) are roaming within its service area. 
\end{samepage}
The User type is determined by the User's preference to protect his location privacy. For a User of type $i \in \mc{A}$, where $\mc{A}$ is the set of User types, his privacy preferences are modeled by $\Pi_i \in [0,1]$, which we call the \emph{privacy factor}. For instance, $\Pi_i=1$ models a User who completely ignores the service provided by the Company in favor of maximizing his privacy. We assume that $\Pi_i$ is entered by the User on his mobile device. 
 
We let $t_i$ denote the connection time that a User of type $i$ chooses. Thus, the location privacy of User type $i$, for connection time $t_i$, is given by $p_i = \frac{T}{t_i} \, \hat{l}$, where we denote $p_i(t_i)$ by $p_i$ for convenience.

\begin{table}[t]
\centering
\caption{List of Symbols}
\label{tab_symbols}
\renewcommand*{\arraystretch}{1.1}
\begin{tabular}{| c | p{6.5cm} |}
\hline Symbol & Description \\ \hline \hline
$\hat{l}$ & Expected localization error\\
\rowcolor{gray!25}
$T$ & User visiting time \\
$\mc{A}$ & Set of User types\\
\rowcolor{gray!25}
$\alpha_i$ & Likelihood of the User being of type $i$\\
$\Pi_i$ & Privacy factor for type $i$ User\\
\rowcolor{gray!25}
$t_i$ & Connection time of type $i$ User \\
$p_i$ & Location privacy of type $i$ User\\
\rowcolor{gray!25}
$\delta$ & Very small value, lower bound of the connection time \\
$S$ & Company's offered service level\\
\rowcolor{gray!25}
$S^*$ & Upper bound of Company's offered service level\\
$\hat{S}$ & Expected service level\\
\rowcolor{gray!25}
$\sigma$ & User experienced service level\\
$\Theta$ & Unit service cost\\
\rowcolor{gray!25}
$\Xi$ & Unit service benefit\\
$\phi_j$ & Probability of the $j$-th service level to be chosen\\
\rowcolor{gray!25}
$\mc{S}_U$ & Set of User's pure strategies\\
$\mc{S}_C$ & Set of Company's pure strategies\\
\rowcolor{gray!25}
$\mu_i$ & Threshold value of the offered expected service level, where the best response strategy of type $i$ User changes\\
\hline
\end{tabular}
\end{table}

\subsection{Offered and Experienced Service Level}
We assume that $\tcompany$ can offer a service level $S \in \mathbb{Z}$, with $0 < S \leq S^*$, to the User. The service level $S$ represents the highest possible additive level of the offered services. We differ the User's \emph{experienced service level} $\expserv$ from $S$, and we assume that $\expserv=S$ if and only if the User stays connected for $t=T$; otherwise, $\expserv<S$. It is easy to see that the highest possible service level that the User can experience equals $S^*$, and it can be obtained only when the Company offers $S^*$ and the User chooses $t=T$.

The experienced service level $\expserv$ is modeled as a linear non-decreasing function. In practice, $\expserv$ is discrete (i.e., the Company gives out a discount or not). Therefore, $\expserv$ gets a connection time $t$ and an offered service level, and it provides an attainable discrete service level as follows
\begin{equation}\label{eq:experienced_level}
 \expserv \dfn \frac{\connectiontime}{\totaltime} \, \offserv.
\end{equation}

\section{Location Privacy Game}
\label{sec_lpg}
\begin{samepage} 
In this section, we define the Location Privacy Game (LPG), which is a 2-player Bayesian Stackelberg game between the Company $C$ and the User $U$. In the LPG, the leader (Company) first commits to his strategy, which is observed by the follower (User).\end{samepage} The Bayesian extension to the Stackelberg game allows us to capture multiple types of followers, where each follower has its own payoff values. We denote the set of User types by $\usertypes$, and the User is of type $i$ with probability $\alpha_i$, decided by Nature \cite{alpcan_network_2010}. 

\subsection{Strategies}
In the LPG, the Company decides upon the offered service level $S$ with knowledge of the probability distribution over the different User types. 
On the other hand, the User wants to consume some of these services while respecting his location privacy preferences. The Company advertises $S$, and the User can observe this and play his best response by choosing an optimal $t$.
The Company wishes that the User will stay connected for as long as possible, and therefore, to be able to construct the entire path that the User has followed; however, each offered service level has a cost, which increases with $S$. This cost is modeled by the monotonically increasing function $\Theta \, S$, where $\Theta$ is a positive constant called the \emph{unit service cost}. 
We also assume that the Company benefits from tracking the User's location, for example, by selling his location data to third parties. We model the Company's benefit as a monotonically increasing function of $t$, which is given by $\Xi \, \frac{1}{p(t)}$, where $\Xi$ is a positive constant called the \emph{unit service benefit}.


The pure strategy choice of the Company is to offer a service level $\offserv$, and we express its strategy set as 
$\strategyset_C:=\{1,\dots,S^*\}$. We also express the set of the User's pure strategies as $\strategyset_U:= [\delta,T]$. 
Note that, for the remainder of this paper, we will denote the $j$-th service level by $S_j$, and the connection time chosen by a User of type $i$ is denoted by $t_i$, as mentioned earlier. 




A player's mixed strategy is a distribution over the set of his pure strategies.
For the Company, the canonical representation of its mixed-strategy space is a discrete probability distribution over the set $\strategyset_C$. We represent a mixed strategy of the Company as an $|\mc{S}_C|$-dimensional vector $\mixedstratcomp$, where $\phi_j$ is the probability of offering the $j$-th service level. In the LPG, we assume that the $\user$ plays only pure strategies, since there always exists a pure strategy that is a best response for the User, as it is also explained in \cite{jain_quality-bounded_2011}.

\subsection{Payoffs}

\subsubsection{Company}

For a given User type $i$ and strategy profile $(\mixedstratcomp, t_i)$, the Company's payoff is
\begin{align}
 \mc{U}_C^{(i)}(\mixedstratcomp,t_i) & \dfn
 \Xi \, \frac{1}{p_i} - \Theta \, \sum\limits_{j\in \mc{S}_C}\phi_j \, S_j \nonumber\\
 &=\frac{\Xi}{T} \, \frac{1}{\hat{l}} \, t_i
 - \Theta \, \sum\limits_{j\in \mc{S}_C}\phi_j \, S_j.
 \label{eq_payoff_company}
\end{align}

This payoff is in the form $\Psi \, t_i - \Theta \, \sum\limits_{j\in \mc{S}_C}\phi_j \, S_j$, where $\Psi, \Theta$, are positive constants, and 
\begin{equation}
  \Psi=\frac{\Xi}{T} \, \frac{1}{\hat{l}} .
\end{equation}


The \emph{overall expected payoff} of the Company is a weighted combination of its expected payoff against all user types. We represent the Users' strategies, one per each type, as an $|\mc{A}|$-dimensional vector $\mf{t}=[t_i]$, where $t_i \in \mc{S}_U$. Then, from Eq.~\eqref{eq_payoff_company}, we have that the Company's overall expected payoff is
\begin{align}
\label{eq_total_payoff_company}
 \mc{U}_C(\mixedstratcomp,\mf{t})&=
 \sum\limits_{i\in \mc{A}} \alpha_i\cdot \mc{U}_C^{(i)}(\mixedstratcomp,t_i) \nonumber \\
 &= \sum\limits_{i\in \mc{A}}\alpha_i\Bigg[\comppayoff\Bigg].
\end{align}

\subsubsection{User}
For a given offered service level $S$ and connection time $t_i$, the $\user$'s payoff is determined by both the achieved privacy and the experienced service level as follows:
\begin{eqnarray}
\label{eq:payoff_user_type}
\mc{U}_{U}^{(i)}(S,t_i) &\dfn&\privacyRandomVariable_i \, p_i + (1-\privacyRandomVariable_i)\,\sigma(t_i,S)\nonumber\\
 &=& \Pi_i \, T \, \hat{l} \, \frac{1}{t_i} + (1-\Pi_i) \, \frac{1}{T} \, S\, t_i, 
\end{eqnarray}
which is in the form $\Psi_1 \, \frac{1}{t_i} + \Psi_2 \, S \, t_i$, where $\Psi_1, \Psi_2$ are positive constants, for a specific User type $i$, and
\begin{equation}
\begin{cases} 
  \Psi_1 = \Pi_i \, T \, \hat{l} \\ 
  \Psi_2 = (1-\Pi_i) \, \frac{1}{T}.
\end{cases}
\end{equation}

Hence, the User's payoff for a mixed strategy $\mixedstratcomp$ of the Company is
\begin{equation}
\label{eq:expPayoffUser}
\mc{U}_{U}^{(i)}(\mixedstratcomp,t_i)=
\Psi_1 \, \frac{1}{t_i} + \Psi_2 \, t_i \, \sum_{j\in \mc{S}_C} \phi_j \, S_j.
\end{equation}
It is easy to see that there is a trade-off between location privacy and experienced service quality level when choosing~$\connectiontime$.
For instance, staying connected for long time leads to high~$\sigma$ but low $\privacy$, and vice versa.

\section{Analysis}
\label{sec_analysis}

In the analysis, our goal will be to find the User's best response and the Company's optimal strategies, which are defined as follows.
\begin{definition}
A User strategy is a \emph{best response} if it
maximizes the User's payoff, taking the Company's offered service level as given.
\end{definition}

The standard solution concept for Stackelberg games is the \emph{Strong Stackelberg Equilibrium} (SSE) \cite{yin_stackelberg_2010}.

\begin{definition}
\label{def_sse_user}
At the Strong Stackelberg Equilibrium (SSE) of the LPG 
\begin{enumerate}
\item for every type $i$, the User of type $i$ plays a best-response $t^*$ to any Company strategy $\mixedstratcomp$, that is,
\[
\mc{U}_{U}^{(i)}(\mixedstratcomp, t^*) \geq
  \mc{U}_{U}^{(i)}(\mixedstratcomp, t), \, \forall \, t \neq t^*;
\] 
\item the Users break ties in favor of the Company, that is, when there are multiple best responses to a Company strategy $\mixedstratcomp$, the Users play the best responses $\mf{t}^*$ that maximize the Company's payoff:
\[
\mc{U}_C(\mixedstratcomp, \mf{t}^*) \geq
  \mc{U}_C(\mixedstratcomp, \mf{t}), \, \forall \, \mf{t} \text{ best response};
\]
\item the Company plays a best-response $\mixedstratcomp^*$, which maximizes its payoff given that the Users' strategies are given by the first two conditions (i.e., Users always play best responses with tie-breaking in favor of the Company):
\[
\mc{U}_C(\mixedstratcomp^*, \mf{t}^*(\mixedstratcomp)) \geq
  \mc{U}_C(\mixedstratcomp, \mf{t}^*(\mixedstratcomp)), \, \forall \, \mixedstratcomp ,
\]
where $\mf{t}^*(\mixedstratcomp)$ denotes the Users best responses with tie-breaking to a Company strategy $\mixedstratcomp$.
\end{enumerate}
\end{definition}
Note that, in our game, the tie-breaking rule has no practical implications, it merely eliminates some pathological mathematical cases where the game would have no equilibrium otherwise.

Since the Company's equilibrium strategies maximize its payoff, given that Users maximize their own payoffs, we will refer to them as optimal strategies for the remainder of the paper. 
\begin{definition}
A Company strategy is \emph{optimal} if it maximizes the Company's payoff given that the User will always play a best-response strategy with tie-breaking in favor of the Company.
\end{definition}

\subsection{Representing the Company's Mixed Strategies}
First, observe that both the Company's and the User's expected payoffs depend on the Company's mixed strategy~$\mixedstratcomp$ only through the expected service level $\sum_{j\in \mc{S}_C} \phi_j \, S_j$. To simplify our analysis, we now introduce $\hat{S}$ to denote the expected service level. For any mixed strategy~$\mixedstratcomp$ of the Company, we can compute the corresponding $\hat{S}$ as $\hat{S} = \sum_{j\in \mc{S}_C} \phi_j \, S_j$. Then, we can express the Company's expected payoff as
\begin{equation}
\label{eq_comp_payoff}
\mc{U}_C(\hat{S}, \mf{t}) = \sum_{i\in \mc{A}}\alpha_i\Big[\Psi \, t_i
 - \Theta \, \hat{S} \Big]
\end{equation}
and the User's expected payoff as
\begin{equation}
\mc{U}_{U}^{(i)}(\hat{S},t_i)=
\Psi_1 \, \frac{1}{t_i} + \Psi_2 \, t_i \, \hat{S} .
\end{equation}
Furthermore, it is also clear that, for any $\hat{S} \in [\min_{j \in \mc{S}_C} S_j, \max_{j \in \mc{S}_C} S_j]$, there exists a mixed strategy $\mixedstratcomp$ for the Company such that $\sum_{j\in \mc{S}_C} \phi_j \, S_j = \hat{S}$. Hence, we can use $\hat{S} \in [\min_{j \in \mc{S}_C} S_j, \max_{j \in \mc{S}_C} S_j]$ to represent the Company's mixed strategies, and the problem of finding an optimal strategy reduces to finding an optimal $\hat{S}$ value.

\subsection{User's Best Response}
In order to find an optimal strategy for the Company, we first have to characterize the Users' best-response strategies.
\begin{lemma}
\label{lem:userBestRespCond}
For any Company strategy $\hat{S}$, the User's best response is either $\delta$ or $T$.
\end{lemma}
\begin{IEEEproof}
The domain of the payoff function $\mc{U}_{U}^{(i)}(\hat{S},t_i)$ is $t_i$ in $[\delta, T]$. Then, we can compute the first derivative of $\mc{U}_{U}^{(i)}(\hat{S},t_i)$ with respect to $t_i$ as $\frac{\partial \mc{U}_{U}^{(i)}}{\partial t_i} = -\Psi_1 \, \frac{1}{t_i^2} + \Psi_2 \, \hat{S}$. Next, we can compute the second derivative of $\mc{U}_{U}^{(i)}(\hat{S},t_i)$ with respect to $t_i$ as $\frac{\partial^2 \mc{U}_{U}^{(i)}}{\partial t_i^2} = 2 \, \Psi_1 \, \frac{1}{t_i^3} + 0 > 0$. Since the second derivative is always positive on $[\delta, t_i]$, we have that the payoff function $\mc{U}_{U}^{(i)}(\hat{S},t_i)$ is a convex function of $t_i$. It follows from the convexity of the function that the maximum payoff is attained at one of the endpoints $\delta$ and $T$. Therefore, the User's best response is either $\delta$ or $T$.
\end{IEEEproof}

\begin{theorem}
\label{thm:userBestRespChar}
If User of type $i$ plays a best-response strategy and breaks ties in favor of the Company, then his strategic choice for a Company strategy $\hat{S}$ is
\begin{itemize}
\item $t_i = \delta$ if $\hat{S} < \mu_i$,
\item $t_i = T$ if $\hat{S} \geq \mu_i$,
\end{itemize}
where
\begin{equation}
\label{eq_mu_i}
	\mu_i = \frac{\Psi_1}{\Psi_2} \, \frac{1}{\delta \, T} .
\end{equation}
\end{theorem}

\begin{figure}[h]
\vspace{-0.3cm}
\centering
\begin{tikzpicture}[x=6cm, y=3cm, font=\small, semithick]
\draw [->] (-0.1, 0) -- (0.95, 0) node [right] {$\hat{S}$};
\draw [->] (0, -0.1) -- (0, 1) node [above] {$t_i$};
\foreach \ytick/\name in {{0.33/\delta}, {0.67/T}}
  \draw (2pt, \ytick) -- (-2pt, \ytick) node [left] {$\name$}; 
\foreach \xtick/\name in {{0.25/$\min_{j \in \mc{S}_C} S_j$}, {0.5/$\mu_i$}, {0.75/$\max_{j \in \mc{S}_C} S_j$}}
  \draw (\xtick, 2pt) -- (\xtick, -2pt) node [below] {\name}; 
\foreach \level/\startx/\endx in {0.33/0.25/0.5, 0.67/0.5/0.75} {
  \draw [shorten >= 1.5pt] (\startx, \level) -- (\endx, \level);
  \fill (\startx, \level) circle [radius=1.5pt];
  \draw (\endx, \level) circle [radius=1.5pt];
}
\fill (0.75, 0.67) circle [radius=1.5pt];
\end{tikzpicture}
\caption{Illustration of the User's best response with tie-breaking as a function of the Company's strategy $\hat{S}$.}
\label{fig:userBestRespFunc}
\end{figure}
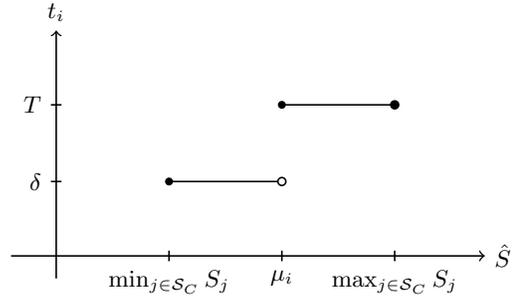

The above theorem basically shows that the User's best response is a non-decreasing right-continuous step function of $\hat{S}$
(see Fig.~\ref{fig:userBestRespFunc} for an illustration).
Note that, if the threshold $\mu_i$ is outside the interval $[\min_{j \in \mc{S}_C} S_j, \max_{j \in \mc{S}_C} S_j]$, then the best response is constant.

\begin{IEEEproof}
From Lemma~\ref{lem:userBestRespCond}, we have that the User's strategic choice is either $\delta$ or $T$. Since the Company's payoff is always an increasing function of $t_i$, the User has to choose $T$ if both $\delta$ and $T$ are best responses, as the User breaks ties in favor of the Company. Hence, it remains to characterize the case when $\delta$ is the only best response. The strategy $\delta$ is a better response than the strategy $T$ if and only if
\begin{eqnarray}
&&\mc{U}_{U}^{(i)}(\hat{S}, \delta) > \mc{U}_{U}^{(i)}(\hat{S}, T)
\Rightarrow \frac{\Psi_1}{\delta} + \Psi_2 \, \delta \, \hat{S} > \frac{\Psi_1}{T} + \Psi_2 \, T \, \hat{S}\nonumber\\
&&\Rightarrow \Psi_1 \, (\frac{1}{\delta}-\frac{1}{T}) > \Psi_2 \, \hat{S} \, (T-\delta) \Rightarrow \Psi_1 \, \frac{T-\delta}{\delta\,T} > \nonumber\\
&&\Psi_2 \, \hat{S} \, (T-\delta) \Rightarrow \hat{S} < \frac{\Psi_1}{\Psi_2} \, \frac{1}{\delta\,T}.
\end{eqnarray}
\end{IEEEproof}

\subsection{Company's Optimal Strategy}
\begin{lemma}
\label{lem:companyOptStratRestr}
Suppose that we are given a set of User strategies $\mf{t} = (t_1, t_2, \dots, t_{\mc{S}_U})$, and the Company's strategy space is limited to $\hat{S}$ values for which $\mf{t}$ is a best response. Then, the Company's payoff is a strictly decreasing function of~$\hat{S}$.
\end{lemma}

\begin{IEEEproof}
We can reformulate the Company's payoff function as
\begin{eqnarray}
\mc{U}_C(\hat{S}) &= \sum_{i\in \mc{A}}\alpha_i\Big[\Psi \, t_i - \Theta \, \hat{S} \Big] \nonumber\\
& = \underbrace{- \Theta}_{< 0} \hat{S} + \underbrace{\sum_{i\in \mc{A}}\alpha_i\Big[\Psi \, t_i \Big]}_{\text{constant}} .
\end{eqnarray}
Hence, on this limited strategy space, the Company's payoff is a strictly decreasing function of $\hat{S}$.
\end{IEEEproof}

\begin{theorem}
The Company's optimal strategy is either $\min_{j \in \mc{S}_C} S_j$ or one of the threshold values $\mu_i$ defined in Theorem~\ref{thm:userBestRespChar}.
\end{theorem}

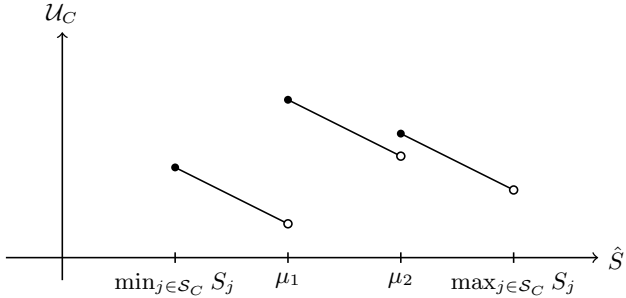
\begin{figure}[h]
\vspace{-0.2cm}
\centering
\begin{tikzpicture}[x=7.5cm, y=3cm, font=\small, semithick]
\draw [->] (-0.1, 0) -- (0.95, 0) node [right] {$\hat{S}$};
\draw [->] (0, -0.1) -- (0, 1) node [above] {$\mc{U}_C$};
\foreach \xtick/\name in {{0.2/$\min_{j \in \mc{S}_C} S_j$}, {0.4/$\mu_1$}, {0.6/$\mu_2$}, {0.8/$\max_{j \in \mc{S}_C} S_j$}}
  \draw (\xtick, 2pt) -- (\xtick, -2pt) node [below] {\name}; 
\foreach \startx/\endx/\starty/\endy in {0.2/0.4/0.4/0.15, 0.4/0.6/0.7/0.45, 0.6/0.8/0.55/0.3} {
  \draw [shorten >= 1.5pt] (\startx, \starty) -- (\endx, \endy);
  \fill (\startx, \starty) circle [radius=1.5pt];
  \draw (\endx, \endy) circle [radius=1.5pt];
}
\end{tikzpicture}
\caption{Illustration of the Company's expected payoff as a function of its strategy $\hat{S}$.
In this figure, the optimal strategy is $\mu_1$.}
\label{fig:companyPayoff}
\vspace{-0.2cm}
\end{figure}

\begin{IEEEproof}
The Users' threshold values $\mu_1, \mu_2, \ldots, \mu_{|\mc{S}_U|}$ divide the Company's strategy space $[\min_{j \in \mc{S}_C} S_j, \max_{j \in \mc{S}_C} S_j]$ into at most $|\mc{S}_U| + 1$ contiguous intervals. From Lemma~\ref{lem:companyOptStratRestr}, we have that the Company's payoff is strictly decreasing on each one of these intervals.
From Lemma~\ref{thm:userBestRespChar}, we have that each of these intervals is left-closed 
(see Fig.~\ref{fig:companyPayoff} for an illustration).
Therefore, the Company's payoff attains its maximum at one of the left endpoints, that is, either at $\min_{j \in \mc{S}_C} S_j$ or at one of the threshold values $\mu_i$.
\end{IEEEproof}

\section{Results}
\label{sec_case_study}
For the purposes of this section, we have used a wireless (IEEE 802.11) localization testbed to derive realistic expected localization error $\hat{l}$ values, which we have then used to derive the payoffs of the Company and the User. We have undertaken simulations to compare the payoffs of different User types. Additionally, we have compared the Bayesian Company strategy with a strategy that assumes that all the Users have the same average $\Pi_i$ value. Finally, we have demonstrated the benefit of our game-theoretic solution as opposed to non-strategic decisions. 

For this case study, we define the set of possible expected service levels as $\{1, 2, \dots, 10\}$. Since LPG is a Stackelberg game, the User is aware of these service levels and he chooses the one that maximizes his payoff. On the other hand, the Company chooses an optimal $\hat{S} \in \{1, 2, \dots, 10\}$.
\begin{samepage}
In our testbed, the measurement stations (MSs) are devices that use the IEEE 802.11 protocol (i.e., WiFi) and their wireless cards are set into monitor mode. We performed practical measurements by using an IEEE 802.11 testbed. 
We have generated Received Signal Strength (RSS) values as inputs to our localization algorithm. 
\end{samepage}
To generate these values, we use the formula \cite{bahillo_hybrid_2010}
\begin{align}\label{eq:RSSGen}
 P_{R_i}=P_0(d_0) - 10 \, n_i \, \log_{10} \frac{d_i}{d_0} + X,~\mbox{where}
\end{align}
\begin{itemize}
 \item $P_{R_i}$ is the received power at station $i$;
 \item $P_0(d_0)$ is a reference power measured at distance $d_0$;
 \item $n_i$ is the path loss exponent, which depends on the environment between User and measurement station $i$;
 \item $d_i$ is the distance between MS $i$ and User's device;
 \item $X$ is a zero-mean log-normal distributed random variable reflecting the flat fading with standard deviation $\epsilon_X$.
\end{itemize}
We have used a Nexus 4 mobile device, which sends 1000 packets per second, and we have selected twelve locations where the User could be. We have taken 1000 measurements at each of these locations, for 4 directions, resulting in 4000 measurements for each location. By averaging these measurements we have derived $n_i=0.75~\forall i$, $d_0=0.7$ meters, $P_0(d_0)=-59$, and $\epsilon_X=1$. 
We use the previously identified values and (\ref{eq:RSSGen}) to simulate and derive a mean localization error when different number of packets are sent by the User device. The latter affect the localization error because of the flat fading $X$. Therefore, we use $1000$ random locations from the interval $[0,10] \times [0,10]$ and locate the User using multilateration \cite[p. 164]{bensky_wireless_2008}. We assume three different values 1000, 500, and 200 for the amount of data sent by a device resulting in the mean localization errors 40.12m, 46.64m, and 58.04m, correspondingly. The errors depend strongly on the environment and obstacles (e.g.,\,moving people, walls) in the propagation path. We also see that multilateration does not perform well at all. However, the performance of this localization system falls out of the scope of this paper.

Following the results of Westin \cite{westin2003social}, we classify the users into the following three categories: Privacy Fundamentalists (PFs); Privacy Unconcerned (PUs); and Privacy Pragmatists (PPs). According to \cite{westin2003social}, PFs ``\emph{reject the consumer-benefit or societal-protection claims for data uses and sought legal-regulatory privacy measure};'' PUs are ``\emph{ready to supply their personal information to business and government and reject what is seen as too much privacy fuss};'' and PPs ``\emph{examine the benefits to them of the data collection and use, want to know the privacy risks and how organizations propose to control those, and then decide whether to trust the organization or seek legal oversight}.'' Therefore, we define the set of User types as $\mc{A}=\{\mbox{PU},\mbox{PP},\mbox{PF}\}$, and we set their corresponding privacy factors $\Pi_i$ as $\{0.2, 0.5, 0.8\}$. We have simulated a scenario where the User's minimal connection time is $\delta=2$, and the unit service benefit $\Xi$ is 50\% higher than the unit service cost $\Theta$.~Note that the above privacy factor values have been chosen for the purpose of evaluating our model and they should not be considered as a recommendation from the literature.~We also recognize that in real-life scenarios we might notice the ``privacy paradox'', according to which people tend to express extreme privacy preferences but act differently, in a rather erratic way.~However, in our work here, we assume that users are rational entities whose actions are consistent with their privacy preferences.

\begin{figure}[t]
\centering
\includegraphics[width=200pt,keepaspectratio]{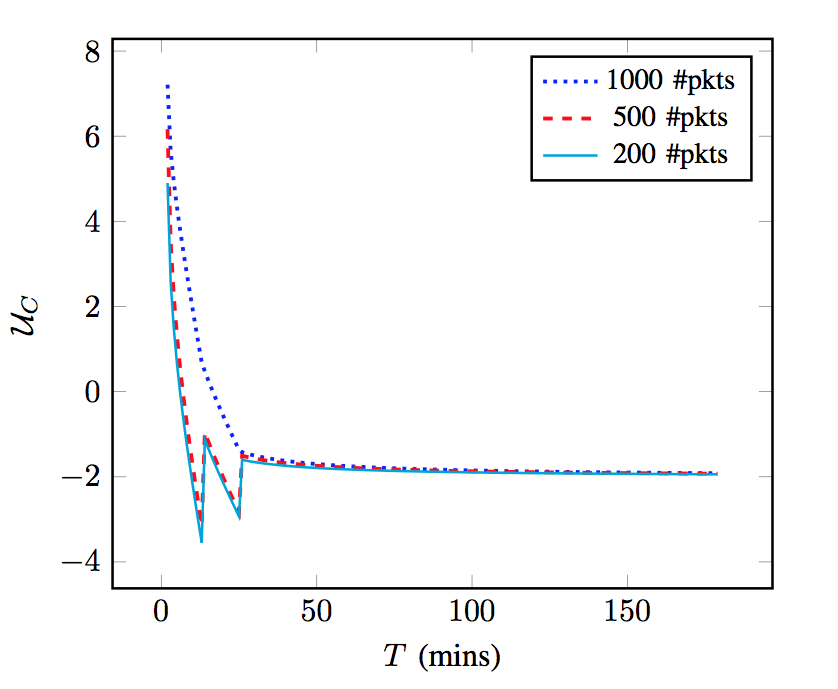}
\vspace{-0.35cm}
\caption{Company's payoff for different number of packets sent by the User device.}
\label{fig_number_of_packets_company}
\end{figure}

Fig.\,\ref{fig_number_of_packets_company} shows the Company's payoff for the different mean localization error values, as discussed previously. We notice that for $\hat{l}=40.12$m, $\mc{U}_C$ becomes negative when $T=17$, and for both $\hat{l}=46.64$m and $\hat{l}=58.04$m, when $T=7$. These low values of total connection time demonstrate the need for an effective localization system, if the Company decides to implement the model discussed in this paper. 

\begin{figure}[t]
\centering
\includegraphics[width=200pt,keepaspectratio]{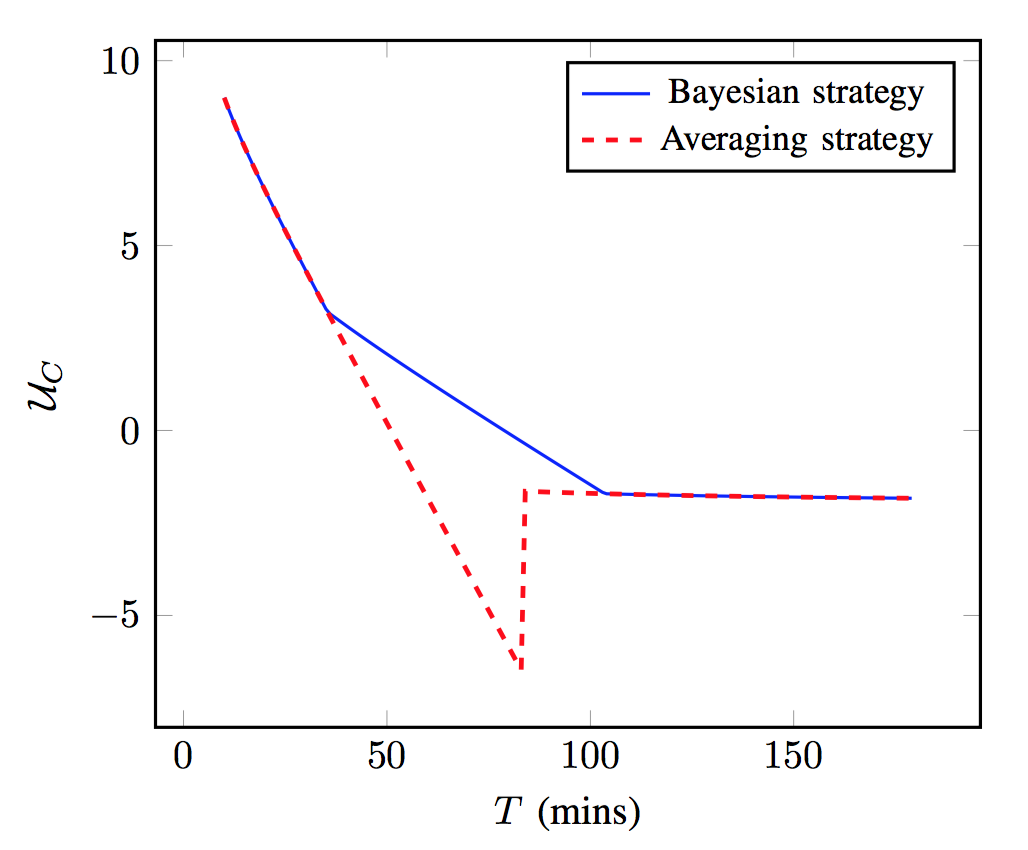}
\vspace{-0.35cm}
\caption{Comparing a Bayesian with an Averaging strategy for the Company.}
\label{fig_bayesian_averaging}
\end{figure}

For the remainder of this paper, we assume that $\hat{l}=2$m. In Fig.\,\ref{fig_bayesian_averaging} we compare the payoffs arising from the optimal Bayesian Company strategy and from the optimal ``Averaging strategy.'' Both strategies are evaluated in the Bayesian model, assuming that User types are uniformly distributed. The former strategy considers the differences between the User types, and as a result, correctly assumes that the privacy factor $\Pi_i$ is drawn from $\{0.2,0.5,0.8\}$ uniformly at random. On the other hand, the Averaging strategy assumes that the users are homogeneous and that the privacy factor always takes its expected value $0.5$ (i.e., assumes a single User type which has the average $\Pi$ value 0.5). This comparison allows us to determine how much the Company can gain from knowing the actual distribution of the User types. For visiting time $T=84$ minutes, the Company's payoff decreases with $T$ for both strategies. However, we notice that the Bayesian Company strategy outperforms the Averaging strategy when $T>36$. Furthermore, the Averaging and Bayesian strategies give negative payoffs for $T>51$, and $T>82$ correspondingly. Given that when negative payoffs are reached the Company must rather decide not to provide any services, the Bayesian strategy gives 31 minutes extra time for the Company to make profit.

\begin{figure}[t]
\centering
\includegraphics[width=200pt,keepaspectratio]{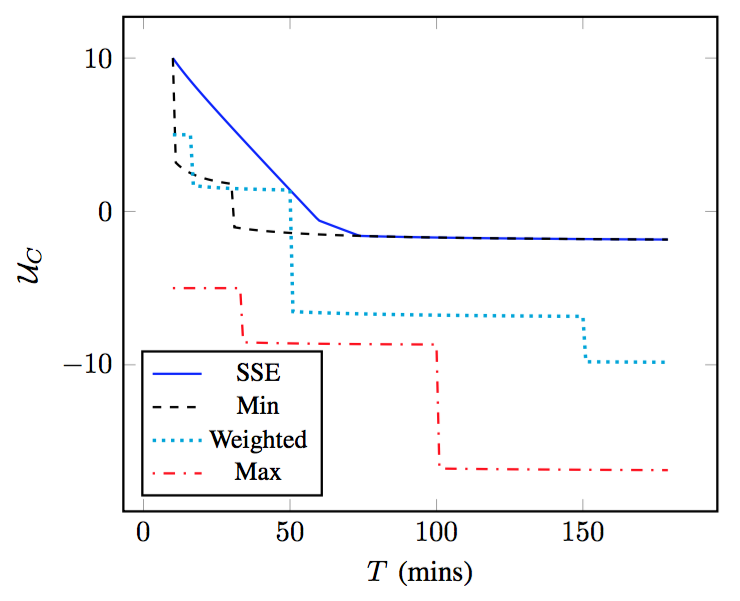}
\vspace{-0.35cm}
\caption{Comparing the payoff of the Company for different non-strategic decisions and the strategy at the Strong Stackelberg Equilibrium (SSE).}
\label{fig_comp_strategies}
\end{figure}

More importantly, in Fig.\,\ref{fig_comp_strategies} we show how the Company benefits from following the Stackelberg strategy as opposed to non-strategic decisions, such as the maximum $\hat{S}$ value 10, the minimum $\hat{S}$ value 1, and also the weighted $\hat{S}$ value. \begin{samepage} The latter is given by first assuming that the Company chooses as expected service levels $[2,5,8]$ when the User privacy factors are $[0.2,0.5,0.8]$. \end{samepage} Secondly, the Company multiplies each $\hat{S}$ value by the probability $\alpha_i$ of a User being of type $i$.

Following the results of \cite{westin2003social}, we have used the probability distribution $\boldsymbol{\alpha} = [\alpha_1,\alpha_2,\alpha_3] = [0.2,0.55,0.25]$ over $\mc{A}$ and, therefore, over $\{0.2,0.5,0.8\}$ for the Bayesian model. We assume that the Company is aware of $\boldsymbol{\alpha}$. Thus, the Company can compute its optimal expected service level, by using Eq.~(\ref{eq_comp_payoff}). It is easy that, for the weighted strategy, given $\boldsymbol{\alpha}$, we have that $\hat{S}=0.2\cdot2 + 0.55\cdot5 + 0.25\cdot8 = \lfloor{5.15}\rfloor = 5$. 

%
%
%

First, we notice that for all of the $\hat{S}$ values, the Company's payoff is a decreasing function of the User's visiting time $T$ in this Bayesian model. More specifically, the results show that if the Company chooses the Max strategy, its payoff becomes negative for $T>10$, while for the Min strategy, the Company can keep providing services for 21 extra minutes ($T=31$), before its payoff becomes negative. This time is improved by 20 minutes when the Company chooses the weighted value, leading to $T=51$ before its payoff becomes negative. The best performance is achieved when the Company chooses the $\hat{S}$ determined by the Strong Stackelberg Equilibrium of the LPG. This allows the Company to make profit (i.e., having positive payoff), for 57 minutes. Although the 6 extra minutes improvement of the Company's payoff per User is not remarkable, we must note that such an improvement leads to significantly higher Company profits when considering a high number of users, as in realistic scenarios. 

It is also worth noting that for $T>51$, the Company's payoff decreases significantly (reaching $-6.529$), when the weighted value is chosen. On the other hand, although the Company's payoff becomes negative for $T>57$, its value remains $-0.0184$  for the rest of the simulated time. This can be useful if we assume that the Company occasionally decides not to stop offering  services immediately after its payoff becomes negative, in favor of its Users.

\begin{figure}[t]
\centering
\includegraphics[width=220pt,keepaspectratio]{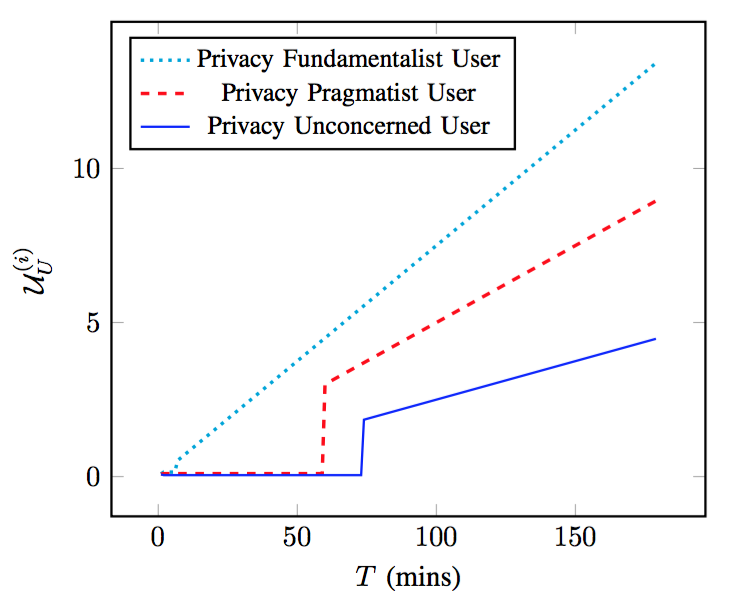}
\vspace{-0.35cm}
\caption{Payoffs of the different User types at the SSE of the LPG.}
\label{fig_users_payoffs}
\end{figure}

Besides investigating the Company's payoff, we have looked into the payoffs of different User types, when the User plays his best response according to Definition \ref{def_sse_user}. In Fig.\,\ref{fig_users_payoffs}, we have plotted these payoffs for the same  parameters $\Theta$, $\Xi$, $\boldsymbol{\alpha}$ as in the above results, and different visiting times $T$. We observe that for a Privacy Fundamentalist (PF), the payoff increases as a function of the visiting time, even from the very first minute. In contrast, the payoff of a Privacy Pragmatist (PP) User equals 0 for visiting times less than 60 minutes. For higher values than this, the User payoff becomes positive taking the value 3.0. Thereafter, for $T>60$ PP's payoff only increases. Finally, the payoff of a Privacy Unconcerned (PU) User, remains 0 for the visiting time values lower than 74. At this point, the payoff becomes 1.85, and thereafter it only increases. However, it remains lower than the PP's payoff for the rest of the time. Note that, both PP's and PU's payoffs are lower than the PF's payoff at all times, highlighting that the latter is the most favored User type in our model.

%
%

\begin{figure}[t]
\centering
\includegraphics[width=215pt,keepaspectratio]{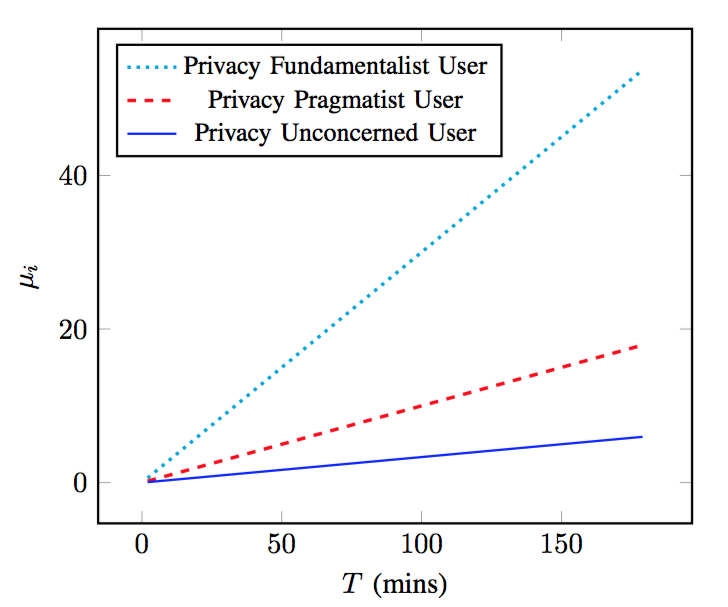}
\vspace{-0.35cm}
\caption{Threshold values, in terms of offered expected service levels, where the User's best response changes, for the different User types.}
\label{fig_users_thresholds}
\end{figure}

Finally, in Fig.~\ref{fig_users_thresholds}, we can see the thresholds for the different User types as a function of the visiting time. As expected, the results show that for all User types, threshold values increase with the visiting time. This means that the higher the visiting time $T$, the higher the expected service level $\hat{S}$ must be for the User to stay connected for $T$, as opposed to remaining connected for a small $\delta$.  This happens because the User is more concerned about his location privacy  for longer visiting periods; therefore, he has to receive a higher $\hat{S}$ in order to consider it worthwhile (i.e., best response) to be connected to the CI of the Company for $T$.

Given that $\hat{S} \in \{1,2,\dots,10\}$, the results show that a PF User will not get connected for more than $\delta$ minutes when the visiting time $T$ exceeds 34 minutes, regardless of the expected service level $\hat{S}$, offered by the Company. Likewise, a PP User considers staying connected to the Company's CI for the whole visiting time, for $T$ values only up to 101 minutes, if the required $\hat{S}$ is offered. Lastly, a PU User can stay connected for the maximum simulation time $T=180$, for the ``right'' $\hat{S}$ value. To have a more clear view on how quickly $\hat{S}$ must increase to satisfy the requirements of the different User types, we have derived the slope of each threshold function for each User type. For this derivation, we have computed the derivative of $\mu_i$ with respect to $T$. Thus, from (\ref{eq_mu_i}), we have $ \frac{\partial \mu_i}{\partial T} = \frac{\partial \frac{\Psi_1}{\Psi_2} \, \frac{1}{\delta \, T}}{\partial T} = \frac{\Psi_1}{\Psi_2} \, \frac{1}{\delta} = \frac{\Pi_i \, \hat{l}}{(1 - \Pi_i) \, \delta}$. From this, we found the following slope values: PU: 0.033, PP: 0.1, and PF: 0.3. We notice that, for the same visiting time, a PF User requires a 3 times higher $\hat{S}$ offered than a PP User, in order to stay connected for $T$, and 9 times higher $\hat{S}$ offered than a PU User.

%
%



\section{Related Work} 
\label{sec_related_work}

In this section we discuss state-of-the-art work at the intersection of game theory and location privacy. A thorough survey related to this has been published by Manshaei et al.~\cite{manshaei_game_2013}. The majority of the papers model two players: the attacker and the user. 

According to \cite{shokri_protecting_2012}, in order to design an optimal privacy-protection mechanism it is crucial to take the \textit{knowledge of an attacker} into consideration. This means, for example, that the adversary is aware of the utilized location protection algorithm and the access profile of the user i.e., the probability distribution describing the user access to Location-based services (LBS) in a certain region. This assumes, that the user contacts the LBS sporadically. 

Shokri et al. \cite{shokri_protecting_2012} provide a framework to methodologically integrate this knowledge by using a zero-sum Bayesian Stackelberg game in order to derive the optimal protection strategy. In their scenario, the user is the leader and the adversary is the follower. They build on the \emph{correctness} metric explained above to measure the users' location privacy. Their game consists of four steps. First, the Nature selects a location $r$ for the user to access the LBS. Second, the user protects his/her location by creating a pseudo-location $r'$ with a function $f$. Third, the attacker observes $r'$ and tries to infer $r$ using the knowledge of $f$ and the access profile of the user resulting in an estimation $\hat{r}$. Finally, the adversary pays an amount $d(r,\hat{r})$ to the user. Here $d(\cdot)$  is a distance function and represents the estimation error of the adversary. The authors derive optimal strategies for both, user and adversary.


Furthermore, Shokri et al. \cite{shokri_collaborative_2011} present a privacy preserving approach relying on user-collaboration. Their solution, called \emph{MobiCrowd}, requires the mobile devices to communicate wirelessly and in a peer-to-peer manner. The mobile devices keep their context information in a buffer, until it expires, and they pass it to other collaborative users seeking such information. This leads to less communication with the service provider because a user contacts the provider only if there are no other users, with the requested information, in range. In this initial work no game theory is used but it is the basis for \cite{santos_collaborative_2011}, where Santos et al. extend their work by analyzing the collaborative behavior of users in MobiCrowd with game-theoretic methods. The two Nash game equilibria, which they have derived, favor mutual cooperation and mutual defection. In a second game they combine game theoretic analysis with an epidemic model to investigate the behavior of more than two users. In this way, they derive the optimal threshold $\alpha_{opt}$ for cooperation that optimizes the payoff of a user. 


Chorppath and Alpcan \cite{chorppath_trading_2013} establish a privacy mechanism-design game between a company and its mobile users. The company offers incentives to the users in order them to report their location with a certain level of accuracy. The authors derive the total budget that a company must invest on providing incentives to obtain a desired minimum level of location accuracy from all the users. 

As far as we know, all above papers make the assumption that users actively report their location in order to use LBSs. They also look into anonymity issues, and they aim to decouple the user identity from his location. However, modern devices do not come with the capability of changing, for instance, their users MAC address, and therefore confusing the attacker about their real identity. 

To the best of our knowledge, our work is the first game-theoretic approach investigating users' strategies in a passive localization environment, where location is derived by raw signal measurements, and the only parameter that the user can control is the amount of connection time. Finally, our work is innovative, as it is the first one to model the economically-rational decision-making of the service provider in conjunction with rational decision-making of the users who wish to protect their location privacy.

\section{Conclusion} 
\label{sec_conclusion}
This paper presents a game-theoretic model, in which a Company incentivizes a User to permit location tracking, by offering ``attractive'' service levels based on the different User types. The User's location is tracked by a passive localization system, which is established and maintained by the Company. We have defined a Stackelberg game, called the Location Privacy Game (LPG), according to which a User selects the amount of time he stays connected (i.e.,\,connection time) to the Company's network (e.g.,\,WiFi), and the Company chooses the level of services that are offered to the User. We have presented theoretical results characterizing the equilibria of the game. Then, we have developed an IEEE 802.11 wireless testbed, which facilitated the computation of different expected localization errors for a User who is equipped with a mobile device. We have used these values in our simulations to demonstrate the superiority of the game-theoretic strategy as opposed to non-strategic methods. More importantly, we have considered different User privacy types, as published by Westin \cite{westin2003social}, and have determined the service level that must be provided by the Company to incentivize the User to stay connected as long as possible to the Company's network. 

Regarding our plans for future work, an interesting and actively explored research direction is developing information theory-based metrics for quantifying user privacy. Therefore modeling user and service provider decision processes using privacy games and by integration of such metrics is another direction that immediately follows. Furthermore, plans include a model extension that will facilitate user privacy within the realm of the Internet-of-Things, where localization capabilities are more often the case than the exception. 

\bibliographystyle{IEEEtran}
\bibliography{paper}
\end{document}